\documentstyle{frank}
\begin{document}
\begin{flushright}
hep-ph/9501401\\
MPI-PhT/95-4\\
January 1995
\end{flushright}

\vskip2cm

\begin{frontmatter}
\title{Light Stops at CDF?}

\author{K. J. Abraham}
\address{{\tt abraham@pixie.udw.ac.za}\\
  Dept.\ of Physics, 
  Univ.\ of Durban-Westville,
  Private Bag X54001,
  Durban 4000,
  South Africa}
\author{Frank Cuypers}
\address{{\tt cuypers@mppmu.mpg.de}\\
  Max-Planck-Institut f\"ur Physik,
  Werner-Heisenberg-Institut,
  F\"ohringer Ring 6, 
  D--80805 M\"unchen, 
  Germany}
\author{Geert Jan van Oldenborgh}
\address{{\tt gj@rulgm0.leidenuniv.nl}\\
  Instituut Lorentz, Rijksuniversiteit Leiden,
  Postbus 9506, NL-2300 RA Leiden,
  Netherlands}
\begin{abstract}
We study the possibility that 
the production and decay of light stop squarks at Tevatron
can mimic top events.
We show 
that this scenario is very unlikely to explain 
the anomalously high top production cross sections 
recently reported by the CDF collaboration.
\end{abstract}
\journal{Physics Letters B}
\end{frontmatter}
\clearpage


\section{Introduction}

Recently the CDF collaboration 
reported the first direct evidence 
for the existence of the top quark $t$ \cite{CDFtop}. 
Although the top mass range 
$m_t = 174 \pm 10{}^{+13}_{-12}$ GeV  
inferred from the reconstruction of 10 events
is compatible 
with precision electro-weak measurements, 
$m_t = 178 \pm 11 {}^{+18}_{-19}$ GeV
\cite{somerecentreview}, 
a number of questions remain unanswered. 
Perhaps the most pressing issue 
is the discrepancy between the measured cross-section 
and its theoretical expectation 
based on the mass range quoted above:
the standard model predicts cross-sections significantly smaller 
than the one observed by CDF. 

This discrepancy has prompted several authors
to invoke the existence of a light stop squark $\tilde{t}$
\cite{stop}.
Indeed,
theoretical expectations lead to a low mass 
for one of the two the supersymmetric partners of the top quark,
if the latter is heavy \cite{er}.
Since a light stop has decay modes 
very similar to those of the top quark, 
this could lead to stop
decays misidentified as top events.
It is the purpose of this letter 
to demonstrate that,
although light stops might well be produced at the Tevatron,
they are very unlikely to explain the anomalously high event rates 
observed by CDF.

We first discuss the production and decay of stops in $p\bar{p}$ collisions, 
establishing under which circumstances 
the signal does in fact resemble that of a top quark.  
Next we investigate the effects of the cuts 
CDF imposed to isolate its top signal 
 from the overwhelming backgrounds.
This allows us to make a comparison with 
the CDF results for a range of values of the top and stop mass,
as well as other relevant supersymmetry parameters.  


\section{Total rates}

In the following we assume the validity 
of the minimal supersymmetric stan\-dard model.
Hence the lightest supersymmetric particle is a neutralino 
$\tilde\chi^{0}_{1}$
which is stable because of $R$-parity conservation
and typically escapes detection.
For a heavy top quark
it is realistic to assume the following hierarchy
in the supersymmetric sector:
\begin{equation}
m_{\tilde\chi^{0}_{1}} < m_{\tilde t} < m_{\mbox{all others}}
\ .\label{hier}
\end{equation}
In this case
the stop can only decay via a virtual chargino 
or a 
(not necessarily virtual)
top quark into
\begin{equation}
\tilde t \to \tilde\chi^0_1 W b
\ .\label{stopdecay}
\end{equation}
This is exactly the same decay signature as the top
$t \to Wb$,
except for the extra missing energy
carried away by the invisible neutralino.
If condition (\ref{hier}) were to be relaxed
and one or more sleptons are lighter than the stop,
there are other decay mechanisms
which compete with (\ref{stopdecay}).
The numerical difference with the results to follow is however not major
and the overall conclusions of this study are not affected.

\begin{figure}[htb]
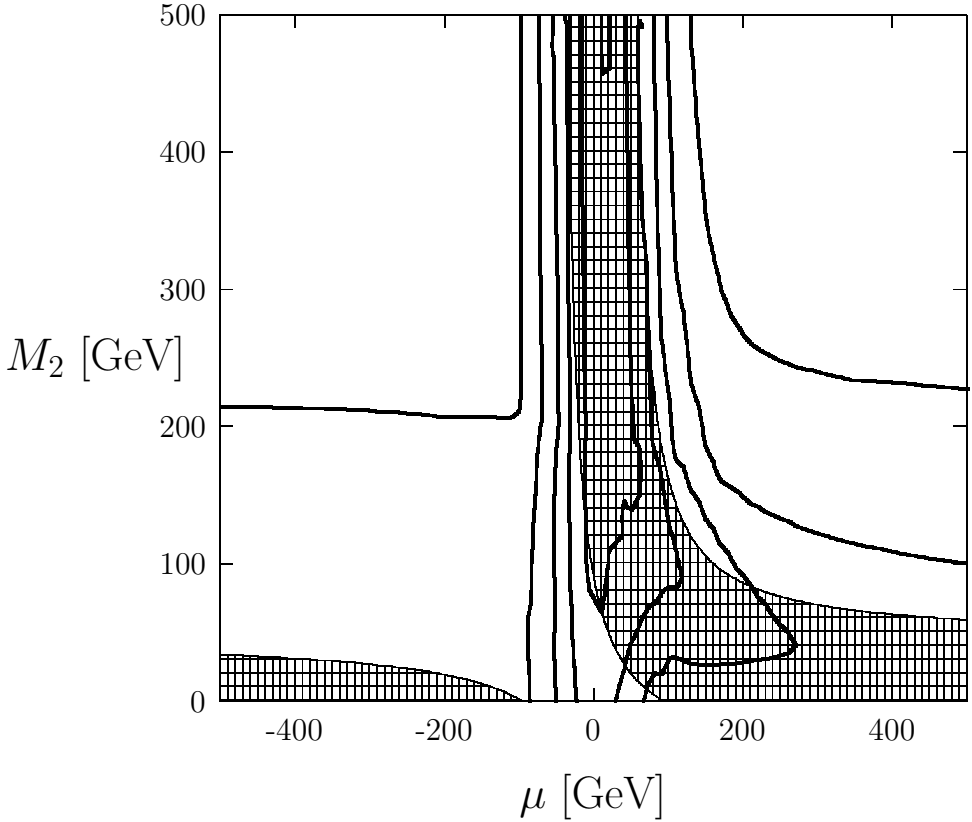

\begin{center}
\setlength{\unitlength}{0.240900pt}
\ifx\plotpoint\undefined\newsavebox{\plotpoint}\fi
\sbox{\plotpoint}{\rule[-0.175pt]{0.350pt}{0.350pt}}%

\end{center}
\caption[]{Contours of the branching ratio $BR(t\to bW)=.5,.6,.7,.8,.9,1$,
  from the innermost to the outermost curves.
  This set of contours has been obtained for 
  $m_t=160$ GeV,
  $m_{\tilde t}=50$ GeV and
  $\tan\beta=1$.
  For larger values of $\tan\beta$ 
  the plot becomes increasingly more symmetric 
  with respect to the axis $\mu=0$.
  The hatched area is excluded by LEP.}
\label{fig4}
\end{figure}

Stop quarks can be obtained from the production and decay of top quarks. 
The rates depend on the branching fractions of the decays
$t \rightarrow \tilde{t} + \tilde\chi^{0}_{i}$
($i=1\dots4$), 
which in turn depends on various supersymmetry parameters.
In practice,
however,
for most of the parameter space not yet excluded by the LEP experiments
this branching ratio does not exceed more than 10--25\%.
This can be seen from Fig.~\ref{fig4},
where we plotted several contours of equal $t \to Wb$ branching fractions
in the plane of the Higgs mixing mass $\mu$
and the supersymmetry breaking mass $M_2$ \cite{susy}.
These contours have been plotted for the heaviest top mass
expected at the $1\sigma$ level 
 from the rates observed at CDF
($m_t=160$ GeV, {\em cf.} next section),
and for the lightest stop mass barely compatible with other experiments.
The branching ratio is thus likely to be closer to one than shown.  
The small branching 
combined with the paucity of top events
mean that this channel can be neglected for stop production 
as compared to direct stop production.  
This also implies that the standard model prediction for top production 
remains largely unaffected by the existence of a lighter stop.  
In the following we will neglect altogether any possible deviation from one 
of the branching ratio $t \to Wb$.

\begin{figure}[htb]
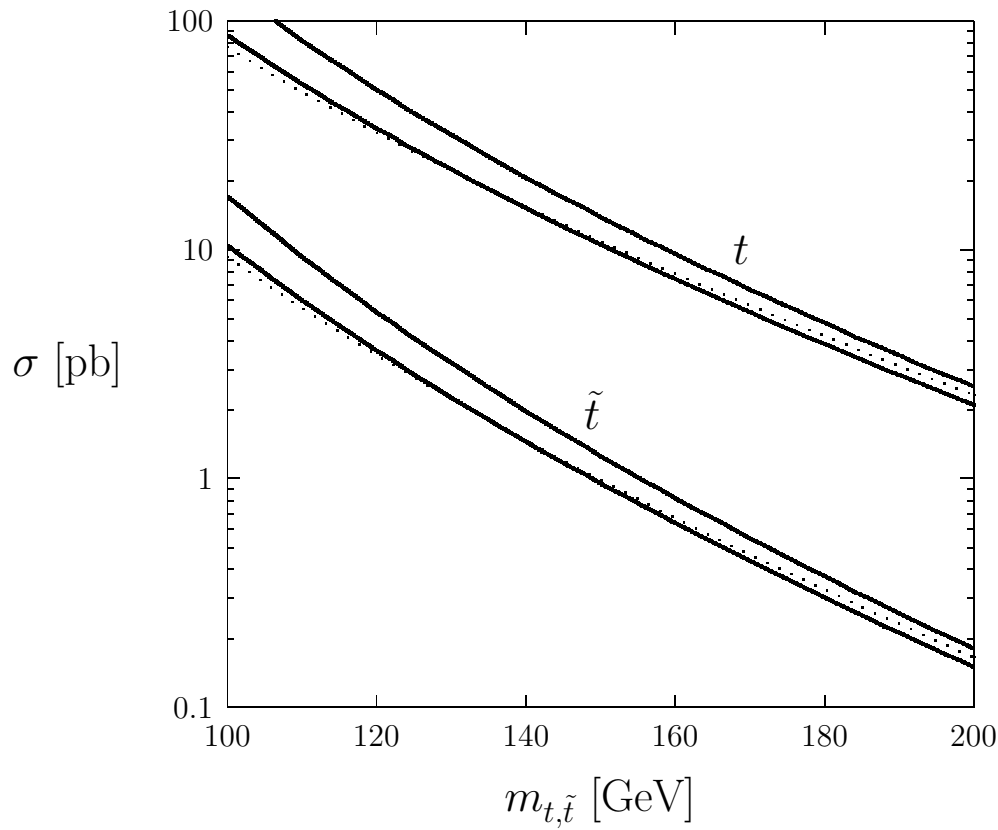

\setlength{\unitlength}{0.240900pt}
\ifx\plotpoint\undefined\newsavebox{\plotpoint}\fi
\sbox{\plotpoint}{\rule[-0.175pt]{0.350pt}{0.350pt}}%

\caption[]{Top and stop production at $\protect\sqrt{s}=1.8$ TeV.
  The thick lines are the next to leading order results of 
  Ref.~\protect\cite{JackWillyTop}.
  The dotted lines are the lowest order result.}
\label{fig1}
\end{figure}

Stops can also be directly pair-produced in hadronic collisions. 
The dominant mechanism at $\sqrt{s} = 1.8$ TeV is 
$ q\bar{q} \to \tilde{t} \tilde{t}$, 
with $gg \rightarrow \tilde{t} \tilde{t}$ giving a small 
additional contribution \cite{DawsonEichtenQuiggSuper}.  
The lowest order cross 
sections at Tevatron 
are shown in Fig.~\ref{fig1} 
along with the corresponding cross section for $t\bar{t}$ production.  
The rates are not sensitive to the structure functions used 
as these are probed at large $x$. 
We have used the GRV LO set \cite{GRV}.
Higher order corrections have been computed 
for top production \cite{JackWillyTop} and are also shown in Fig.~\ref{fig1}.
The large uncertainty is due to the choice of scale.
In the absence of a similar 
${\cal O}(\alpha_s^3)$
calculation for the squark sector,
we have assumed the same K-factors to hold for stop production.  
It appears that the supersymmetric channel could make a sizable contribution 
if the mass of the stop is about 50 GeV less than that of the top.  

\section{Observable rates}

The CDF group concentrates on two types of top signals,
according to two of the three possible decay modes of the $W$ bosons
which emerge from the decay of the top:
\begin{itemize}
\item
  dilepton: 
  $t\bar{t}\to\ell^+\nu_\ell\ell'^-\bar{\nu}_{\ell'} b\bar{b}$;
\item
  lepton+jets:
  $t\bar{t}\to\ell^\pm\stackrel{\scriptscriptstyle(-)}{\nu}_\ell q\bar{q}' 
        b\bar{b}$.
\end{itemize}

The cuts required by CDF to isolate the signal 
are described in detail in Refs~\cite{CDFtop}.  
To obtain an estimate of the influence of these cuts on 
the observability of the stop signal,
we have performed a rough simulation of top and stop production and decay, 
assuming isotropic boosted decays 
and neglecting all fragmentation, hadronization and detector effects 
({\em i.e.}, jets were represented by a single parton).  
Using the simulation 
we obtain efficiencies which are higher than the ones quoted by CDF 
by a factor 1.2 (lepton+jets) to 2 (dilepton)
for $m_t=160$ GeV.
The difference is due to the isolation and jet cuts, 
which cannot be implemented well at the parton level.  
The effect of these cuts, 
and hence the difference between our simulation and the CDF results, 
increases for lower masses.

The expected number of events $n$ is computed as
\begin{equation}
    n(m_t,m_{\tilde{t}}) = {\cal L} \!\!
        \sum_{\shortstack{\tiny dileptons\\\tiny lepton+jets}} \!\!
        \sigma_t(m_t) \varepsilon_t^{\rm CDF}(m_t)
      + \sigma_{\tilde t}(m_{\tilde{t}}) \left( 
              \varepsilon_{\tilde{t}}(m_{\tilde{t}}) 
        \over \varepsilon_t(m_{\tilde{t}}) \right) 
        \varepsilon_t^{\rm CDF}(m_{\tilde{t}})
\ ,\label{events}
\end{equation}
where ${\cal L}=19.3$ pb is the integrated luminosity
analyzed by CDF.
The ratio of efficiencies $\varepsilon_{\tilde t}/\varepsilon_t$ 
has been computed using our parton level Monte Carlo simulation,
and should be fairly insensitive to the isolation and jet cuts.
For masses exceeding 100 GeV
the efficiencies $\varepsilon_t^{\rm CDF}$ 
are taken from Refs~\cite{CDFtop}.
For lower masses no efficiencies are quoted
and we have used our computed efficiencies
corrected by the extrapolation of the factors mentioned above.

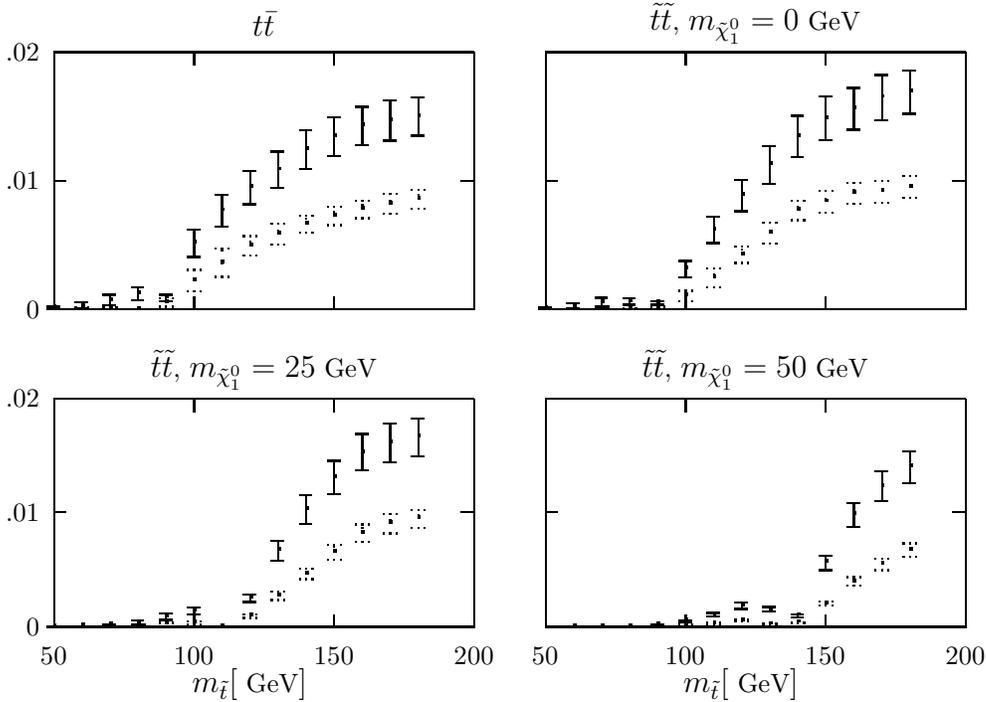
\begin{figure}[htb]
\begin{center}
\hspace*{-3.5cm}
\setlength{\unitlength}{0.240900pt}
\ifx\plotpoint\undefined\newsavebox{\plotpoint}\fi
\begin{picture}(900,540)(0,0)
\font\gnuplot=cmr10 at 10pt
\gnuplot
\sbox{\plotpoint}{\rule[-0.200pt]{0.400pt}{0.400pt}}%
\put(176.0,68.0){\rule[-0.200pt]{158.994pt}{0.400pt}}
\put(176.0,68.0){\rule[-0.200pt]{4.818pt}{0.400pt}}
\put(154,68){\makebox(0,0)[r]{0}}
\put(816.0,68.0){\rule[-0.200pt]{4.818pt}{0.400pt}}
\put(176.0,270.0){\rule[-0.200pt]{4.818pt}{0.400pt}}
\put(154,270){\makebox(0,0)[r]{.01}}
\put(816.0,270.0){\rule[-0.200pt]{4.818pt}{0.400pt}}
\put(176.0,472.0){\rule[-0.200pt]{4.818pt}{0.400pt}}
\put(154,472){\makebox(0,0)[r]{.02}}
\put(816.0,472.0){\rule[-0.200pt]{4.818pt}{0.400pt}}
\put(176.0,68.0){\rule[-0.200pt]{0.400pt}{4.818pt}}
\put(176.0,452.0){\rule[-0.200pt]{0.400pt}{4.818pt}}
\put(396.0,68.0){\rule[-0.200pt]{0.400pt}{4.818pt}}
\put(396.0,452.0){\rule[-0.200pt]{0.400pt}{4.818pt}}
\put(616.0,68.0){\rule[-0.200pt]{0.400pt}{4.818pt}}
\put(616.0,452.0){\rule[-0.200pt]{0.400pt}{4.818pt}}
\put(836.0,68.0){\rule[-0.200pt]{0.400pt}{4.818pt}}
\put(836.0,452.0){\rule[-0.200pt]{0.400pt}{4.818pt}}
\put(176.0,68.0){\rule[-0.200pt]{158.994pt}{0.400pt}}
\put(836.0,68.0){\rule[-0.200pt]{0.400pt}{97.324pt}}
\put(176.0,472.0){\rule[-0.200pt]{158.994pt}{0.400pt}}
\put(506,517){\makebox(0,0){$t\bar{t}$}}
\put(176.0,68.0){\rule[-0.200pt]{0.400pt}{97.324pt}}
\put(220,68){\rule{1pt}{1pt}}
\put(264,69){\rule{1pt}{1pt}}
\put(308,68){\rule{1pt}{1pt}}
\put(352,79){\rule{1pt}{1pt}}
\put(396,113){\rule{1pt}{1pt}}
\put(440,141){\rule{1pt}{1pt}}
\put(484,168){\rule{1pt}{1pt}}
\put(528,186){\rule{1pt}{1pt}}
\put(572,202){\rule{1pt}{1pt}}
\put(616,214){\rule{1pt}{1pt}}
\put(660,225){\rule{1pt}{1pt}}
\put(704,233){\rule{1pt}{1pt}}
\put(748,241){\rule{1pt}{1pt}}
\put(220.00,68.00){\usebox{\plotpoint}}
\put(220,71){\usebox{\plotpoint}}
\put(210.00,68.00){\usebox{\plotpoint}}
\put(230,68){\usebox{\plotpoint}}
\put(210.00,71.00){\usebox{\plotpoint}}
\put(230,71){\usebox{\plotpoint}}
\put(264.00,68.00){\usebox{\plotpoint}}
\put(264,71){\usebox{\plotpoint}}
\put(254.00,68.00){\usebox{\plotpoint}}
\put(274,68){\usebox{\plotpoint}}
\put(254.00,71.00){\usebox{\plotpoint}}
\put(274,71){\usebox{\plotpoint}}
\put(308,68){\usebox{\plotpoint}}
\put(308,68){\usebox{\plotpoint}}
\put(298.00,68.00){\usebox{\plotpoint}}
\put(318,68){\usebox{\plotpoint}}
\put(298.00,68.00){\usebox{\plotpoint}}
\put(318,68){\usebox{\plotpoint}}
\put(352.00,72.00){\usebox{\plotpoint}}
\put(352,85){\usebox{\plotpoint}}
\put(342.00,72.00){\usebox{\plotpoint}}
\put(362,72){\usebox{\plotpoint}}
\put(342.00,85.00){\usebox{\plotpoint}}
\put(362,85){\usebox{\plotpoint}}
\multiput(396,96)(0.000,20.756){2}{\usebox{\plotpoint}}
\put(396,130){\usebox{\plotpoint}}
\put(386.00,96.00){\usebox{\plotpoint}}
\put(406,96){\usebox{\plotpoint}}
\put(386.00,130.00){\usebox{\plotpoint}}
\put(406,130){\usebox{\plotpoint}}
\multiput(440,119)(0.000,20.756){3}{\usebox{\plotpoint}}
\put(440,163){\usebox{\plotpoint}}
\put(430.00,119.00){\usebox{\plotpoint}}
\put(450,119){\usebox{\plotpoint}}
\put(430.00,163.00){\usebox{\plotpoint}}
\put(450,163){\usebox{\plotpoint}}
\multiput(484,153)(0.000,20.756){2}{\usebox{\plotpoint}}
\put(484,183){\usebox{\plotpoint}}
\put(474.00,153.00){\usebox{\plotpoint}}
\put(494,153){\usebox{\plotpoint}}
\put(474.00,183.00){\usebox{\plotpoint}}
\put(494,183){\usebox{\plotpoint}}
\multiput(528,170)(0.000,20.756){2}{\usebox{\plotpoint}}
\put(528,202){\usebox{\plotpoint}}
\put(518.00,170.00){\usebox{\plotpoint}}
\put(538,170){\usebox{\plotpoint}}
\put(518.00,202.00){\usebox{\plotpoint}}
\put(538,202){\usebox{\plotpoint}}
\multiput(572,189)(0.000,20.756){2}{\usebox{\plotpoint}}
\put(572,215){\usebox{\plotpoint}}
\put(562.00,189.00){\usebox{\plotpoint}}
\put(582,189){\usebox{\plotpoint}}
\put(562.00,215.00){\usebox{\plotpoint}}
\put(582,215){\usebox{\plotpoint}}
\multiput(616,200)(0.000,20.756){2}{\usebox{\plotpoint}}
\put(616,229){\usebox{\plotpoint}}
\put(606.00,200.00){\usebox{\plotpoint}}
\put(626,200){\usebox{\plotpoint}}
\put(606.00,229.00){\usebox{\plotpoint}}
\put(626,229){\usebox{\plotpoint}}
\multiput(660,211)(0.000,20.756){2}{\usebox{\plotpoint}}
\put(660,239){\usebox{\plotpoint}}
\put(650.00,211.00){\usebox{\plotpoint}}
\put(670,211){\usebox{\plotpoint}}
\put(650.00,239.00){\usebox{\plotpoint}}
\put(670,239){\usebox{\plotpoint}}
\multiput(704,218)(0.000,20.756){2}{\usebox{\plotpoint}}
\put(704,249){\usebox{\plotpoint}}
\put(694.00,218.00){\usebox{\plotpoint}}
\put(714,218){\usebox{\plotpoint}}
\put(694.00,249.00){\usebox{\plotpoint}}
\put(714,249){\usebox{\plotpoint}}
\multiput(748,226)(0.000,20.756){2}{\usebox{\plotpoint}}
\put(748,256){\usebox{\plotpoint}}
\put(738.00,226.00){\usebox{\plotpoint}}
\put(758,226){\usebox{\plotpoint}}
\put(738.00,256.00){\usebox{\plotpoint}}
\put(758,256){\usebox{\plotpoint}}
\put(176,70){\rule{1pt}{1pt}}
\put(220,74){\rule{1pt}{1pt}}
\put(264,82){\rule{1pt}{1pt}}
\put(308,92){\rule{1pt}{1pt}}
\put(352,86){\rule{1pt}{1pt}}
\put(396,172){\rule{1pt}{1pt}}
\put(440,223){\rule{1pt}{1pt}}
\put(484,259){\rule{1pt}{1pt}}
\put(528,287){\rule{1pt}{1pt}}
\put(572,319){\rule{1pt}{1pt}}
\put(616,339){\rule{1pt}{1pt}}
\put(660,356){\rule{1pt}{1pt}}
\put(704,365){\rule{1pt}{1pt}}
\put(748,371){\rule{1pt}{1pt}}
\put(176.0,68.0){\rule[-0.200pt]{0.400pt}{0.964pt}}
\put(166.0,68.0){\rule[-0.200pt]{4.818pt}{0.400pt}}
\put(166.0,72.0){\rule[-0.200pt]{4.818pt}{0.400pt}}
\put(220.0,69.0){\rule[-0.200pt]{0.400pt}{2.409pt}}
\put(210.0,69.0){\rule[-0.200pt]{4.818pt}{0.400pt}}
\put(210.0,79.0){\rule[-0.200pt]{4.818pt}{0.400pt}}
\put(264.0,74.0){\rule[-0.200pt]{0.400pt}{4.095pt}}
\put(254.0,74.0){\rule[-0.200pt]{4.818pt}{0.400pt}}
\put(254.0,91.0){\rule[-0.200pt]{4.818pt}{0.400pt}}
\put(308.0,82.0){\rule[-0.200pt]{0.400pt}{5.059pt}}
\put(298.0,82.0){\rule[-0.200pt]{4.818pt}{0.400pt}}
\put(298.0,103.0){\rule[-0.200pt]{4.818pt}{0.400pt}}
\put(352.0,81.0){\rule[-0.200pt]{0.400pt}{2.409pt}}
\put(342.0,81.0){\rule[-0.200pt]{4.818pt}{0.400pt}}
\put(342.0,91.0){\rule[-0.200pt]{4.818pt}{0.400pt}}
\put(396.0,150.0){\rule[-0.200pt]{0.400pt}{10.359pt}}
\put(386.0,150.0){\rule[-0.200pt]{4.818pt}{0.400pt}}
\put(386.0,193.0){\rule[-0.200pt]{4.818pt}{0.400pt}}
\put(440.0,198.0){\rule[-0.200pt]{0.400pt}{12.045pt}}
\put(430.0,198.0){\rule[-0.200pt]{4.818pt}{0.400pt}}
\put(430.0,248.0){\rule[-0.200pt]{4.818pt}{0.400pt}}
\put(484.0,233.0){\rule[-0.200pt]{0.400pt}{12.527pt}}
\put(474.0,233.0){\rule[-0.200pt]{4.818pt}{0.400pt}}
\put(474.0,285.0){\rule[-0.200pt]{4.818pt}{0.400pt}}
\put(528.0,259.0){\rule[-0.200pt]{0.400pt}{13.731pt}}
\put(518.0,259.0){\rule[-0.200pt]{4.818pt}{0.400pt}}
\put(518.0,316.0){\rule[-0.200pt]{4.818pt}{0.400pt}}
\put(572.0,289.0){\rule[-0.200pt]{0.400pt}{14.454pt}}
\put(562.0,289.0){\rule[-0.200pt]{4.818pt}{0.400pt}}
\put(562.0,349.0){\rule[-0.200pt]{4.818pt}{0.400pt}}
\put(616.0,309.0){\rule[-0.200pt]{0.400pt}{14.695pt}}
\put(606.0,309.0){\rule[-0.200pt]{4.818pt}{0.400pt}}
\put(606.0,370.0){\rule[-0.200pt]{4.818pt}{0.400pt}}
\put(660.0,326.0){\rule[-0.200pt]{0.400pt}{14.454pt}}
\put(650.0,326.0){\rule[-0.200pt]{4.818pt}{0.400pt}}
\put(650.0,386.0){\rule[-0.200pt]{4.818pt}{0.400pt}}
\put(704.0,333.0){\rule[-0.200pt]{0.400pt}{15.177pt}}
\put(694.0,333.0){\rule[-0.200pt]{4.818pt}{0.400pt}}
\put(694.0,396.0){\rule[-0.200pt]{4.818pt}{0.400pt}}
\put(748.0,341.0){\rule[-0.200pt]{0.400pt}{14.454pt}}
\put(738.0,341.0){\rule[-0.200pt]{4.818pt}{0.400pt}}
\put(738.0,401.0){\rule[-0.200pt]{4.818pt}{0.400pt}}
\end{picture}
\hspace*{-1.5cm}
\setlength{\unitlength}{0.240900pt}
\ifx\plotpoint\undefined\newsavebox{\plotpoint}\fi
\begin{picture}(900,540)(0,0)
\font\gnuplot=cmr10 at 10pt
\gnuplot
\sbox{\plotpoint}{\rule[-0.200pt]{0.400pt}{0.400pt}}%
\put(176.0,68.0){\rule[-0.200pt]{158.994pt}{0.400pt}}
\put(176.0,68.0){\rule[-0.200pt]{4.818pt}{0.400pt}}
\put(816.0,68.0){\rule[-0.200pt]{4.818pt}{0.400pt}}
\put(176.0,270.0){\rule[-0.200pt]{4.818pt}{0.400pt}}
\put(816.0,270.0){\rule[-0.200pt]{4.818pt}{0.400pt}}
\put(176.0,472.0){\rule[-0.200pt]{4.818pt}{0.400pt}}
\put(816.0,472.0){\rule[-0.200pt]{4.818pt}{0.400pt}}
\put(176.0,68.0){\rule[-0.200pt]{0.400pt}{4.818pt}}
\put(176.0,452.0){\rule[-0.200pt]{0.400pt}{4.818pt}}
\put(396.0,68.0){\rule[-0.200pt]{0.400pt}{4.818pt}}
\put(396.0,452.0){\rule[-0.200pt]{0.400pt}{4.818pt}}
\put(616.0,68.0){\rule[-0.200pt]{0.400pt}{4.818pt}}
\put(616.0,452.0){\rule[-0.200pt]{0.400pt}{4.818pt}}
\put(836.0,68.0){\rule[-0.200pt]{0.400pt}{4.818pt}}
\put(836.0,452.0){\rule[-0.200pt]{0.400pt}{4.818pt}}
\put(176.0,68.0){\rule[-0.200pt]{158.994pt}{0.400pt}}
\put(836.0,68.0){\rule[-0.200pt]{0.400pt}{97.324pt}}
\put(176.0,472.0){\rule[-0.200pt]{158.994pt}{0.400pt}}
\put(506,517){\makebox(0,0){$\tilde{t}\tilde{t}$, 
  $m_{\tilde{\chi}^0_1} = 0$ GeV}}
\put(176.0,68.0){\rule[-0.200pt]{0.400pt}{97.324pt}}
\put(220,68){\rule{1pt}{1pt}}
\put(264,69){\rule{1pt}{1pt}}
\put(308,69){\rule{1pt}{1pt}}
\put(352,71){\rule{1pt}{1pt}}
\put(396,89){\rule{1pt}{1pt}}
\put(440,117){\rule{1pt}{1pt}}
\put(484,154){\rule{1pt}{1pt}}
\put(528,188){\rule{1pt}{1pt}}
\put(572,224){\rule{1pt}{1pt}}
\put(616,237){\rule{1pt}{1pt}}
\put(660,250){\rule{1pt}{1pt}}
\put(704,253){\rule{1pt}{1pt}}
\put(748,260){\rule{1pt}{1pt}}
\put(220,68){\usebox{\plotpoint}}
\put(220,68){\usebox{\plotpoint}}
\put(210.00,68.00){\usebox{\plotpoint}}
\put(230,68){\usebox{\plotpoint}}
\put(210.00,68.00){\usebox{\plotpoint}}
\put(230,68){\usebox{\plotpoint}}
\put(264.00,68.00){\usebox{\plotpoint}}
\put(264,72){\usebox{\plotpoint}}
\put(254.00,68.00){\usebox{\plotpoint}}
\put(274,68){\usebox{\plotpoint}}
\put(254.00,72.00){\usebox{\plotpoint}}
\put(274,72){\usebox{\plotpoint}}
\put(308.00,68.00){\usebox{\plotpoint}}
\put(308,70){\usebox{\plotpoint}}
\put(298.00,68.00){\usebox{\plotpoint}}
\put(318,68){\usebox{\plotpoint}}
\put(298.00,70.00){\usebox{\plotpoint}}
\put(318,70){\usebox{\plotpoint}}
\put(352.00,69.00){\usebox{\plotpoint}}
\put(352,73){\usebox{\plotpoint}}
\put(342.00,69.00){\usebox{\plotpoint}}
\put(362,69){\usebox{\plotpoint}}
\put(342.00,73.00){\usebox{\plotpoint}}
\put(362,73){\usebox{\plotpoint}}
\put(396.00,81.00){\usebox{\plotpoint}}
\put(396,97){\usebox{\plotpoint}}
\put(386.00,81.00){\usebox{\plotpoint}}
\put(406,81){\usebox{\plotpoint}}
\put(386.00,97.00){\usebox{\plotpoint}}
\put(406,97){\usebox{\plotpoint}}
\multiput(440,102)(0.000,20.756){2}{\usebox{\plotpoint}}
\put(440,132){\usebox{\plotpoint}}
\put(430.00,102.00){\usebox{\plotpoint}}
\put(450,102){\usebox{\plotpoint}}
\put(430.00,132.00){\usebox{\plotpoint}}
\put(450,132){\usebox{\plotpoint}}
\multiput(484,141)(0.000,20.756){2}{\usebox{\plotpoint}}
\put(484,167){\usebox{\plotpoint}}
\put(474.00,141.00){\usebox{\plotpoint}}
\put(494,141){\usebox{\plotpoint}}
\put(474.00,167.00){\usebox{\plotpoint}}
\put(494,167){\usebox{\plotpoint}}
\multiput(528,171)(0.000,20.756){2}{\usebox{\plotpoint}}
\put(528,204){\usebox{\plotpoint}}
\put(518.00,171.00){\usebox{\plotpoint}}
\put(538,171){\usebox{\plotpoint}}
\put(518.00,204.00){\usebox{\plotpoint}}
\put(538,204){\usebox{\plotpoint}}
\multiput(572,208)(0.000,20.756){2}{\usebox{\plotpoint}}
\put(572,239){\usebox{\plotpoint}}
\put(562.00,208.00){\usebox{\plotpoint}}
\put(582,208){\usebox{\plotpoint}}
\put(562.00,239.00){\usebox{\plotpoint}}
\put(582,239){\usebox{\plotpoint}}
\multiput(616,220)(0.000,20.756){2}{\usebox{\plotpoint}}
\put(616,254){\usebox{\plotpoint}}
\put(606.00,220.00){\usebox{\plotpoint}}
\put(626,220){\usebox{\plotpoint}}
\put(606.00,254.00){\usebox{\plotpoint}}
\put(626,254){\usebox{\plotpoint}}
\multiput(660,234)(0.000,20.756){2}{\usebox{\plotpoint}}
\put(660,267){\usebox{\plotpoint}}
\put(650.00,234.00){\usebox{\plotpoint}}
\put(670,234){\usebox{\plotpoint}}
\put(650.00,267.00){\usebox{\plotpoint}}
\put(670,267){\usebox{\plotpoint}}
\multiput(704,235)(0.000,20.756){2}{\usebox{\plotpoint}}
\put(704,270){\usebox{\plotpoint}}
\put(694.00,235.00){\usebox{\plotpoint}}
\put(714,235){\usebox{\plotpoint}}
\put(694.00,270.00){\usebox{\plotpoint}}
\put(714,270){\usebox{\plotpoint}}
\multiput(748,243)(0.000,20.756){2}{\usebox{\plotpoint}}
\put(748,277){\usebox{\plotpoint}}
\put(738.00,243.00){\usebox{\plotpoint}}
\put(758,243){\usebox{\plotpoint}}
\put(738.00,277.00){\usebox{\plotpoint}}
\put(758,277){\usebox{\plotpoint}}
\put(176,69){\rule{1pt}{1pt}}
\put(220,73){\rule{1pt}{1pt}}
\put(264,79){\rule{1pt}{1pt}}
\put(308,80){\rule{1pt}{1pt}}
\put(352,78){\rule{1pt}{1pt}}
\put(396,131){\rule{1pt}{1pt}}
\put(440,193){\rule{1pt}{1pt}}
\put(484,247){\rule{1pt}{1pt}}
\put(528,295){\rule{1pt}{1pt}}
\put(572,340){\rule{1pt}{1pt}}
\put(616,368){\rule{1pt}{1pt}}
\put(660,383){\rule{1pt}{1pt}}
\put(704,401){\rule{1pt}{1pt}}
\put(748,409){\rule{1pt}{1pt}}
\put(176.0,68.0){\rule[-0.200pt]{0.400pt}{0.723pt}}
\put(166.0,68.0){\rule[-0.200pt]{4.818pt}{0.400pt}}
\put(166.0,71.0){\rule[-0.200pt]{4.818pt}{0.400pt}}
\put(220.0,69.0){\rule[-0.200pt]{0.400pt}{1.927pt}}
\put(210.0,69.0){\rule[-0.200pt]{4.818pt}{0.400pt}}
\put(210.0,77.0){\rule[-0.200pt]{4.818pt}{0.400pt}}
\put(264.0,72.0){\rule[-0.200pt]{0.400pt}{3.373pt}}
\put(254.0,72.0){\rule[-0.200pt]{4.818pt}{0.400pt}}
\put(254.0,86.0){\rule[-0.200pt]{4.818pt}{0.400pt}}
\put(308.0,75.0){\rule[-0.200pt]{0.400pt}{2.409pt}}
\put(298.0,75.0){\rule[-0.200pt]{4.818pt}{0.400pt}}
\put(298.0,85.0){\rule[-0.200pt]{4.818pt}{0.400pt}}
\put(352.0,75.0){\rule[-0.200pt]{0.400pt}{1.445pt}}
\put(342.0,75.0){\rule[-0.200pt]{4.818pt}{0.400pt}}
\put(342.0,81.0){\rule[-0.200pt]{4.818pt}{0.400pt}}
\put(396.0,118.0){\rule[-0.200pt]{0.400pt}{6.263pt}}
\put(386.0,118.0){\rule[-0.200pt]{4.818pt}{0.400pt}}
\put(386.0,144.0){\rule[-0.200pt]{4.818pt}{0.400pt}}
\put(440.0,172.0){\rule[-0.200pt]{0.400pt}{9.877pt}}
\put(430.0,172.0){\rule[-0.200pt]{4.818pt}{0.400pt}}
\put(430.0,213.0){\rule[-0.200pt]{4.818pt}{0.400pt}}
\put(484.0,222.0){\rule[-0.200pt]{0.400pt}{11.804pt}}
\put(474.0,222.0){\rule[-0.200pt]{4.818pt}{0.400pt}}
\put(474.0,271.0){\rule[-0.200pt]{4.818pt}{0.400pt}}
\put(528.0,265.0){\rule[-0.200pt]{0.400pt}{14.213pt}}
\put(518.0,265.0){\rule[-0.200pt]{4.818pt}{0.400pt}}
\put(518.0,324.0){\rule[-0.200pt]{4.818pt}{0.400pt}}
\put(572.0,307.0){\rule[-0.200pt]{0.400pt}{15.658pt}}
\put(562.0,307.0){\rule[-0.200pt]{4.818pt}{0.400pt}}
\put(562.0,372.0){\rule[-0.200pt]{4.818pt}{0.400pt}}
\put(616.0,334.0){\rule[-0.200pt]{0.400pt}{16.381pt}}
\put(606.0,334.0){\rule[-0.200pt]{4.818pt}{0.400pt}}
\put(606.0,402.0){\rule[-0.200pt]{4.818pt}{0.400pt}}
\put(660.0,350.0){\rule[-0.200pt]{0.400pt}{15.899pt}}
\put(650.0,350.0){\rule[-0.200pt]{4.818pt}{0.400pt}}
\put(650.0,416.0){\rule[-0.200pt]{4.818pt}{0.400pt}}
\put(704.0,365.0){\rule[-0.200pt]{0.400pt}{17.104pt}}
\put(694.0,365.0){\rule[-0.200pt]{4.818pt}{0.400pt}}
\put(694.0,436.0){\rule[-0.200pt]{4.818pt}{0.400pt}}
\put(748.0,375.0){\rule[-0.200pt]{0.400pt}{16.381pt}}
\put(738.0,375.0){\rule[-0.200pt]{4.818pt}{0.400pt}}
\put(738.0,443.0){\rule[-0.200pt]{4.818pt}{0.400pt}}
\end{picture}
\hspace*{-3.0cm}
\\
\hspace*{-3.5cm}
\setlength{\unitlength}{0.240900pt}
\ifx\plotpoint\undefined\newsavebox{\plotpoint}\fi
\begin{picture}(900,540)(0,0)
\font\gnuplot=cmr10 at 10pt
\gnuplot
\sbox{\plotpoint}{\rule[-0.200pt]{0.400pt}{0.400pt}}%
\put(176.0,113.0){\rule[-0.200pt]{158.994pt}{0.400pt}}
\put(176.0,113.0){\rule[-0.200pt]{4.818pt}{0.400pt}}
\put(154,113){\makebox(0,0)[r]{0}}
\put(816.0,113.0){\rule[-0.200pt]{4.818pt}{0.400pt}}
\put(176.0,293.0){\rule[-0.200pt]{4.818pt}{0.400pt}}
\put(154,293){\makebox(0,0)[r]{.01}}
\put(816.0,293.0){\rule[-0.200pt]{4.818pt}{0.400pt}}
\put(176.0,472.0){\rule[-0.200pt]{4.818pt}{0.400pt}}
\put(154,472){\makebox(0,0)[r]{.02}}
\put(816.0,472.0){\rule[-0.200pt]{4.818pt}{0.400pt}}
\put(176.0,113.0){\rule[-0.200pt]{0.400pt}{4.818pt}}
\put(176,68){\makebox(0,0){50}}
\put(176.0,452.0){\rule[-0.200pt]{0.400pt}{4.818pt}}
\put(396.0,113.0){\rule[-0.200pt]{0.400pt}{4.818pt}}
\put(396,68){\makebox(0,0){100}}
\put(396.0,452.0){\rule[-0.200pt]{0.400pt}{4.818pt}}
\put(616.0,113.0){\rule[-0.200pt]{0.400pt}{4.818pt}}
\put(616,68){\makebox(0,0){150}}
\put(616.0,452.0){\rule[-0.200pt]{0.400pt}{4.818pt}}
\put(836.0,113.0){\rule[-0.200pt]{0.400pt}{4.818pt}}
\put(836,68){\makebox(0,0){200}}
\put(836.0,452.0){\rule[-0.200pt]{0.400pt}{4.818pt}}
\put(176.0,113.0){\rule[-0.200pt]{158.994pt}{0.400pt}}
\put(836.0,113.0){\rule[-0.200pt]{0.400pt}{86.483pt}}
\put(176.0,472.0){\rule[-0.200pt]{158.994pt}{0.400pt}}
\put(484,23){\makebox(0,0){$m_{\tilde{t}} [\makebox{ GeV}]$}}
\put(506,517){\makebox(0,0){$\tilde{t}\tilde{t}$, 
  $m_{\tilde{\chi}^0_1} = 25$ GeV}}
\put(176.0,113.0){\rule[-0.200pt]{0.400pt}{86.483pt}}
\put(220,113){\rule{1pt}{1pt}}
\put(264,113){\rule{1pt}{1pt}}
\put(308,114){\rule{1pt}{1pt}}
\put(352,117){\rule{1pt}{1pt}}
\put(396,119){\rule{1pt}{1pt}}
\put(440,113){\rule{1pt}{1pt}}
\put(484,130){\rule{1pt}{1pt}}
\put(528,161){\rule{1pt}{1pt}}
\put(572,196){\rule{1pt}{1pt}}
\put(616,230){\rule{1pt}{1pt}}
\put(660,260){\rule{1pt}{1pt}}
\put(704,275){\rule{1pt}{1pt}}
\put(748,283){\rule{1pt}{1pt}}
\put(220,113){\usebox{\plotpoint}}
\put(220,113){\usebox{\plotpoint}}
\put(210.00,113.00){\usebox{\plotpoint}}
\put(230,113){\usebox{\plotpoint}}
\put(210.00,113.00){\usebox{\plotpoint}}
\put(230,113){\usebox{\plotpoint}}
\put(264.00,113.00){\usebox{\plotpoint}}
\put(264,114){\usebox{\plotpoint}}
\put(254.00,113.00){\usebox{\plotpoint}}
\put(274,113){\usebox{\plotpoint}}
\put(254.00,114.00){\usebox{\plotpoint}}
\put(274,114){\usebox{\plotpoint}}
\put(308.00,113.00){\usebox{\plotpoint}}
\put(308,116){\usebox{\plotpoint}}
\put(298.00,113.00){\usebox{\plotpoint}}
\put(318,113){\usebox{\plotpoint}}
\put(298.00,116.00){\usebox{\plotpoint}}
\put(318,116){\usebox{\plotpoint}}
\put(352.00,114.00){\usebox{\plotpoint}}
\put(352,119){\usebox{\plotpoint}}
\put(342.00,114.00){\usebox{\plotpoint}}
\put(362,114){\usebox{\plotpoint}}
\put(342.00,119.00){\usebox{\plotpoint}}
\put(362,119){\usebox{\plotpoint}}
\put(396.00,117.00){\usebox{\plotpoint}}
\put(396,121){\usebox{\plotpoint}}
\put(386.00,117.00){\usebox{\plotpoint}}
\put(406,117){\usebox{\plotpoint}}
\put(386.00,121.00){\usebox{\plotpoint}}
\put(406,121){\usebox{\plotpoint}}
\put(440,113){\usebox{\plotpoint}}
\put(440,113){\usebox{\plotpoint}}
\put(430.00,113.00){\usebox{\plotpoint}}
\put(450,113){\usebox{\plotpoint}}
\put(430.00,113.00){\usebox{\plotpoint}}
\put(450,113){\usebox{\plotpoint}}
\put(484.00,127.00){\usebox{\plotpoint}}
\put(484,132){\usebox{\plotpoint}}
\put(474.00,127.00){\usebox{\plotpoint}}
\put(494,127){\usebox{\plotpoint}}
\put(474.00,132.00){\usebox{\plotpoint}}
\put(494,132){\usebox{\plotpoint}}
\put(528.00,155.00){\usebox{\plotpoint}}
\put(528,168){\usebox{\plotpoint}}
\put(518.00,155.00){\usebox{\plotpoint}}
\put(538,155){\usebox{\plotpoint}}
\put(518.00,168.00){\usebox{\plotpoint}}
\put(538,168){\usebox{\plotpoint}}
\put(572.00,188.00){\usebox{\plotpoint}}
\put(572,204){\usebox{\plotpoint}}
\put(562.00,188.00){\usebox{\plotpoint}}
\put(582,188){\usebox{\plotpoint}}
\put(562.00,204.00){\usebox{\plotpoint}}
\put(582,204){\usebox{\plotpoint}}
\multiput(616,218)(0.000,20.756){2}{\usebox{\plotpoint}}
\put(616,242){\usebox{\plotpoint}}
\put(606.00,218.00){\usebox{\plotpoint}}
\put(626,218){\usebox{\plotpoint}}
\put(606.00,242.00){\usebox{\plotpoint}}
\put(626,242){\usebox{\plotpoint}}
\multiput(660,247)(0.000,20.756){2}{\usebox{\plotpoint}}
\put(660,274){\usebox{\plotpoint}}
\put(650.00,247.00){\usebox{\plotpoint}}
\put(670,247){\usebox{\plotpoint}}
\put(650.00,274.00){\usebox{\plotpoint}}
\put(670,274){\usebox{\plotpoint}}
\multiput(704,260)(0.000,20.756){2}{\usebox{\plotpoint}}
\put(704,290){\usebox{\plotpoint}}
\put(694.00,260.00){\usebox{\plotpoint}}
\put(714,260){\usebox{\plotpoint}}
\put(694.00,290.00){\usebox{\plotpoint}}
\put(714,290){\usebox{\plotpoint}}
\multiput(748,268)(0.000,20.756){2}{\usebox{\plotpoint}}
\put(748,297){\usebox{\plotpoint}}
\put(738.00,268.00){\usebox{\plotpoint}}
\put(758,268){\usebox{\plotpoint}}
\put(738.00,297.00){\usebox{\plotpoint}}
\put(758,297){\usebox{\plotpoint}}
\put(176,113){\rule{1pt}{1pt}}
\put(220,114){\rule{1pt}{1pt}}
\put(264,116){\rule{1pt}{1pt}}
\put(308,120){\rule{1pt}{1pt}}
\put(352,129){\rule{1pt}{1pt}}
\put(396,138){\rule{1pt}{1pt}}
\put(440,113){\rule{1pt}{1pt}}
\put(484,158){\rule{1pt}{1pt}}
\put(528,233){\rule{1pt}{1pt}}
\put(572,298){\rule{1pt}{1pt}}
\put(616,348){\rule{1pt}{1pt}}
\put(660,387){\rule{1pt}{1pt}}
\put(704,402){\rule{1pt}{1pt}}
\put(748,411){\rule{1pt}{1pt}}
\put(176,113){\usebox{\plotpoint}}
\put(166.0,113.0){\rule[-0.200pt]{4.818pt}{0.400pt}}
\put(166.0,113.0){\rule[-0.200pt]{4.818pt}{0.400pt}}
\put(220.0,113.0){\usebox{\plotpoint}}
\put(210.0,113.0){\rule[-0.200pt]{4.818pt}{0.400pt}}
\put(210.0,114.0){\rule[-0.200pt]{4.818pt}{0.400pt}}
\put(264.0,114.0){\rule[-0.200pt]{0.400pt}{0.723pt}}
\put(254.0,114.0){\rule[-0.200pt]{4.818pt}{0.400pt}}
\put(254.0,117.0){\rule[-0.200pt]{4.818pt}{0.400pt}}
\put(308.0,117.0){\rule[-0.200pt]{0.400pt}{1.445pt}}
\put(298.0,117.0){\rule[-0.200pt]{4.818pt}{0.400pt}}
\put(298.0,123.0){\rule[-0.200pt]{4.818pt}{0.400pt}}
\put(352.0,124.0){\rule[-0.200pt]{0.400pt}{2.409pt}}
\put(342.0,124.0){\rule[-0.200pt]{4.818pt}{0.400pt}}
\put(342.0,134.0){\rule[-0.200pt]{4.818pt}{0.400pt}}
\put(396.0,133.0){\rule[-0.200pt]{0.400pt}{2.409pt}}
\put(386.0,133.0){\rule[-0.200pt]{4.818pt}{0.400pt}}
\put(386.0,143.0){\rule[-0.200pt]{4.818pt}{0.400pt}}
\put(440,113){\usebox{\plotpoint}}
\put(430.0,113.0){\rule[-0.200pt]{4.818pt}{0.400pt}}
\put(430.0,113.0){\rule[-0.200pt]{4.818pt}{0.400pt}}
\put(484.0,152.0){\rule[-0.200pt]{0.400pt}{2.891pt}}
\put(474.0,152.0){\rule[-0.200pt]{4.818pt}{0.400pt}}
\put(474.0,164.0){\rule[-0.200pt]{4.818pt}{0.400pt}}
\put(528.0,217.0){\rule[-0.200pt]{0.400pt}{7.468pt}}
\put(518.0,217.0){\rule[-0.200pt]{4.818pt}{0.400pt}}
\put(518.0,248.0){\rule[-0.200pt]{4.818pt}{0.400pt}}
\put(572.0,275.0){\rule[-0.200pt]{0.400pt}{10.840pt}}
\put(562.0,275.0){\rule[-0.200pt]{4.818pt}{0.400pt}}
\put(562.0,320.0){\rule[-0.200pt]{4.818pt}{0.400pt}}
\put(616.0,321.0){\rule[-0.200pt]{0.400pt}{12.768pt}}
\put(606.0,321.0){\rule[-0.200pt]{4.818pt}{0.400pt}}
\put(606.0,374.0){\rule[-0.200pt]{4.818pt}{0.400pt}}
\put(660.0,359.0){\rule[-0.200pt]{0.400pt}{13.731pt}}
\put(650.0,359.0){\rule[-0.200pt]{4.818pt}{0.400pt}}
\put(650.0,416.0){\rule[-0.200pt]{4.818pt}{0.400pt}}
\put(704.0,372.0){\rule[-0.200pt]{0.400pt}{14.695pt}}
\put(694.0,372.0){\rule[-0.200pt]{4.818pt}{0.400pt}}
\put(694.0,433.0){\rule[-0.200pt]{4.818pt}{0.400pt}}
\put(748.0,381.0){\rule[-0.200pt]{0.400pt}{14.213pt}}
\put(738.0,381.0){\rule[-0.200pt]{4.818pt}{0.400pt}}
\put(738.0,440.0){\rule[-0.200pt]{4.818pt}{0.400pt}}
\end{picture}
\hspace*{-1.5cm}
\setlength{\unitlength}{0.240900pt}
\ifx\plotpoint\undefined\newsavebox{\plotpoint}\fi
\begin{picture}(900,540)(0,0)
\font\gnuplot=cmr10 at 10pt
\gnuplot
\sbox{\plotpoint}{\rule[-0.200pt]{0.400pt}{0.400pt}}%
\put(176.0,113.0){\rule[-0.200pt]{158.994pt}{0.400pt}}
\put(176.0,113.0){\rule[-0.200pt]{4.818pt}{0.400pt}}
\put(816.0,113.0){\rule[-0.200pt]{4.818pt}{0.400pt}}
\put(176.0,293.0){\rule[-0.200pt]{4.818pt}{0.400pt}}
\put(816.0,293.0){\rule[-0.200pt]{4.818pt}{0.400pt}}
\put(176.0,472.0){\rule[-0.200pt]{4.818pt}{0.400pt}}
\put(816.0,472.0){\rule[-0.200pt]{4.818pt}{0.400pt}}
\put(176.0,113.0){\rule[-0.200pt]{0.400pt}{4.818pt}}
\put(176,68){\makebox(0,0){50}}
\put(176.0,452.0){\rule[-0.200pt]{0.400pt}{4.818pt}}
\put(396.0,113.0){\rule[-0.200pt]{0.400pt}{4.818pt}}
\put(396,68){\makebox(0,0){100}}
\put(396.0,452.0){\rule[-0.200pt]{0.400pt}{4.818pt}}
\put(616.0,113.0){\rule[-0.200pt]{0.400pt}{4.818pt}}
\put(616,68){\makebox(0,0){150}}
\put(616.0,452.0){\rule[-0.200pt]{0.400pt}{4.818pt}}
\put(836.0,113.0){\rule[-0.200pt]{0.400pt}{4.818pt}}
\put(836,68){\makebox(0,0){200}}
\put(836.0,452.0){\rule[-0.200pt]{0.400pt}{4.818pt}}
\put(176.0,113.0){\rule[-0.200pt]{158.994pt}{0.400pt}}
\put(836.0,113.0){\rule[-0.200pt]{0.400pt}{86.483pt}}
\put(176.0,472.0){\rule[-0.200pt]{158.994pt}{0.400pt}}
\put(484,23){\makebox(0,0){$m_{\tilde{t}} [\makebox{ GeV}]$}}
\put(506,517){\makebox(0,0){$\tilde{t}\tilde{t}$, 
  $m_{\tilde{\chi}^0_1} = 50$ GeV}}
\put(176.0,113.0){\rule[-0.200pt]{0.400pt}{86.483pt}}
\put(220,113){\rule{1pt}{1pt}}
\put(264,113){\rule{1pt}{1pt}}
\put(308,113){\rule{1pt}{1pt}}
\put(352,113){\rule{1pt}{1pt}}
\put(396,115){\rule{1pt}{1pt}}
\put(440,118){\rule{1pt}{1pt}}
\put(484,122){\rule{1pt}{1pt}}
\put(528,117){\rule{1pt}{1pt}}
\put(572,120){\rule{1pt}{1pt}}
\put(616,149){\rule{1pt}{1pt}}
\put(660,184){\rule{1pt}{1pt}}
\put(704,211){\rule{1pt}{1pt}}
\put(748,234){\rule{1pt}{1pt}}
\put(220,113){\usebox{\plotpoint}}
\put(220,113){\usebox{\plotpoint}}
\put(210.00,113.00){\usebox{\plotpoint}}
\put(230,113){\usebox{\plotpoint}}
\put(210.00,113.00){\usebox{\plotpoint}}
\put(230,113){\usebox{\plotpoint}}
\put(264,113){\usebox{\plotpoint}}
\put(264,113){\usebox{\plotpoint}}
\put(254.00,113.00){\usebox{\plotpoint}}
\put(274,113){\usebox{\plotpoint}}
\put(254.00,113.00){\usebox{\plotpoint}}
\put(274,113){\usebox{\plotpoint}}
\put(308,113){\usebox{\plotpoint}}
\put(308,113){\usebox{\plotpoint}}
\put(298.00,113.00){\usebox{\plotpoint}}
\put(318,113){\usebox{\plotpoint}}
\put(298.00,113.00){\usebox{\plotpoint}}
\put(318,113){\usebox{\plotpoint}}
\put(352.00,113.00){\usebox{\plotpoint}}
\put(352,114){\usebox{\plotpoint}}
\put(342.00,113.00){\usebox{\plotpoint}}
\put(362,113){\usebox{\plotpoint}}
\put(342.00,114.00){\usebox{\plotpoint}}
\put(362,114){\usebox{\plotpoint}}
\put(396.00,114.00){\usebox{\plotpoint}}
\put(396,116){\usebox{\plotpoint}}
\put(386.00,114.00){\usebox{\plotpoint}}
\put(406,114){\usebox{\plotpoint}}
\put(386.00,116.00){\usebox{\plotpoint}}
\put(406,116){\usebox{\plotpoint}}
\put(440.00,117.00){\usebox{\plotpoint}}
\put(440,120){\usebox{\plotpoint}}
\put(430.00,117.00){\usebox{\plotpoint}}
\put(450,117){\usebox{\plotpoint}}
\put(430.00,120.00){\usebox{\plotpoint}}
\put(450,120){\usebox{\plotpoint}}
\put(484.00,121.00){\usebox{\plotpoint}}
\put(484,124){\usebox{\plotpoint}}
\put(474.00,121.00){\usebox{\plotpoint}}
\put(494,121){\usebox{\plotpoint}}
\put(474.00,124.00){\usebox{\plotpoint}}
\put(494,124){\usebox{\plotpoint}}
\put(528.00,117.00){\usebox{\plotpoint}}
\put(528,118){\usebox{\plotpoint}}
\put(518.00,117.00){\usebox{\plotpoint}}
\put(538,117){\usebox{\plotpoint}}
\put(518.00,118.00){\usebox{\plotpoint}}
\put(538,118){\usebox{\plotpoint}}
\put(572.00,119.00){\usebox{\plotpoint}}
\put(572,120){\usebox{\plotpoint}}
\put(562.00,119.00){\usebox{\plotpoint}}
\put(582,119){\usebox{\plotpoint}}
\put(562.00,120.00){\usebox{\plotpoint}}
\put(582,120){\usebox{\plotpoint}}
\put(616.00,146.00){\usebox{\plotpoint}}
\put(616,153){\usebox{\plotpoint}}
\put(606.00,146.00){\usebox{\plotpoint}}
\put(626,146){\usebox{\plotpoint}}
\put(606.00,153.00){\usebox{\plotpoint}}
\put(626,153){\usebox{\plotpoint}}
\put(660.00,178.00){\usebox{\plotpoint}}
\put(660,191){\usebox{\plotpoint}}
\put(650.00,178.00){\usebox{\plotpoint}}
\put(670,178){\usebox{\plotpoint}}
\put(650.00,191.00){\usebox{\plotpoint}}
\put(670,191){\usebox{\plotpoint}}
\put(704.00,202.00){\usebox{\plotpoint}}
\put(704,220){\usebox{\plotpoint}}
\put(694.00,202.00){\usebox{\plotpoint}}
\put(714,202){\usebox{\plotpoint}}
\put(694.00,220.00){\usebox{\plotpoint}}
\put(714,220){\usebox{\plotpoint}}
\multiput(748,223)(0.000,20.756){2}{\usebox{\plotpoint}}
\put(748,244){\usebox{\plotpoint}}
\put(738.00,223.00){\usebox{\plotpoint}}
\put(758,223){\usebox{\plotpoint}}
\put(738.00,244.00){\usebox{\plotpoint}}
\put(758,244){\usebox{\plotpoint}}
\put(220,113){\rule{1pt}{1pt}}
\put(264,113){\rule{1pt}{1pt}}
\put(308,113){\rule{1pt}{1pt}}
\put(352,116){\rule{1pt}{1pt}}
\put(396,122){\rule{1pt}{1pt}}
\put(440,132){\rule{1pt}{1pt}}
\put(484,146){\rule{1pt}{1pt}}
\put(528,140){\rule{1pt}{1pt}}
\put(572,131){\rule{1pt}{1pt}}
\put(616,214){\rule{1pt}{1pt}}
\put(660,289){\rule{1pt}{1pt}}
\put(704,334){\rule{1pt}{1pt}}
\put(748,364){\rule{1pt}{1pt}}
\put(220,113){\usebox{\plotpoint}}
\put(210.0,113.0){\rule[-0.200pt]{4.818pt}{0.400pt}}
\put(210.0,113.0){\rule[-0.200pt]{4.818pt}{0.400pt}}
\put(264,113){\usebox{\plotpoint}}
\put(254.0,113.0){\rule[-0.200pt]{4.818pt}{0.400pt}}
\put(254.0,113.0){\rule[-0.200pt]{4.818pt}{0.400pt}}
\put(308.0,113.0){\usebox{\plotpoint}}
\put(298.0,113.0){\rule[-0.200pt]{4.818pt}{0.400pt}}
\put(298.0,114.0){\rule[-0.200pt]{4.818pt}{0.400pt}}
\put(352.0,115.0){\usebox{\plotpoint}}
\put(342.0,115.0){\rule[-0.200pt]{4.818pt}{0.400pt}}
\put(342.0,116.0){\rule[-0.200pt]{4.818pt}{0.400pt}}
\put(396.0,120.0){\rule[-0.200pt]{0.400pt}{0.723pt}}
\put(386.0,120.0){\rule[-0.200pt]{4.818pt}{0.400pt}}
\put(386.0,123.0){\rule[-0.200pt]{4.818pt}{0.400pt}}
\put(440.0,129.0){\rule[-0.200pt]{0.400pt}{1.445pt}}
\put(430.0,129.0){\rule[-0.200pt]{4.818pt}{0.400pt}}
\put(430.0,135.0){\rule[-0.200pt]{4.818pt}{0.400pt}}
\put(484.0,141.0){\rule[-0.200pt]{0.400pt}{2.409pt}}
\put(474.0,141.0){\rule[-0.200pt]{4.818pt}{0.400pt}}
\put(474.0,151.0){\rule[-0.200pt]{4.818pt}{0.400pt}}
\put(528.0,137.0){\rule[-0.200pt]{0.400pt}{1.686pt}}
\put(518.0,137.0){\rule[-0.200pt]{4.818pt}{0.400pt}}
\put(518.0,144.0){\rule[-0.200pt]{4.818pt}{0.400pt}}
\put(572.0,128.0){\rule[-0.200pt]{0.400pt}{1.204pt}}
\put(562.0,128.0){\rule[-0.200pt]{4.818pt}{0.400pt}}
\put(562.0,133.0){\rule[-0.200pt]{4.818pt}{0.400pt}}
\put(616.0,202.0){\rule[-0.200pt]{0.400pt}{5.541pt}}
\put(606.0,202.0){\rule[-0.200pt]{4.818pt}{0.400pt}}
\put(606.0,225.0){\rule[-0.200pt]{4.818pt}{0.400pt}}
\put(660.0,270.0){\rule[-0.200pt]{0.400pt}{8.913pt}}
\put(650.0,270.0){\rule[-0.200pt]{4.818pt}{0.400pt}}
\put(650.0,307.0){\rule[-0.200pt]{4.818pt}{0.400pt}}
\put(704.0,311.0){\rule[-0.200pt]{0.400pt}{11.322pt}}
\put(694.0,311.0){\rule[-0.200pt]{4.818pt}{0.400pt}}
\put(694.0,358.0){\rule[-0.200pt]{4.818pt}{0.400pt}}
\put(748.0,339.0){\rule[-0.200pt]{0.400pt}{12.045pt}}
\put(738.0,339.0){\rule[-0.200pt]{4.818pt}{0.400pt}}
\put(738.0,389.0){\rule[-0.200pt]{4.818pt}{0.400pt}}
\end{picture}
\hspace*{-3.0cm}
\end{center}
\caption[]{Estimated efficiencies times branching ratios for the detection of 
top and stop pairs using the CDF top quark cuts.  The upper solid points 
are for the dilepton channel, the lower dotted ones the dilepton channel}
\label{fig2}
\end{figure}

We have plotted in Figs~\ref{fig2} the efficiencies 
for the top and stop signals,
the latter for three different neutralino masses.
The sharp drops at the 
$m_t = m_W + m_b$ and
$m_{\tilde t} = m_{\tilde\chi^0_1} + m_W + m_b$ 
thresholds
are due to a cut of 85 GeV imposed by CDF 
on the scalar sum of observed transverse momenta
(both in the dilepton and lepton+jets channels).  
Indeed,
as the structure functions are quite steep 
most (s)top pairs are produced close to threshold.  
The decay $W$ bosons and $b$ quarks 
are then produced at rest in the (s)top rest frame, 
which is hardly boosted,
and the decay products of the $W$ 
are unable to pass the stringent transverse momentum cut.

\section{Comparison with the CDF results}

\begin{figure}[htb]
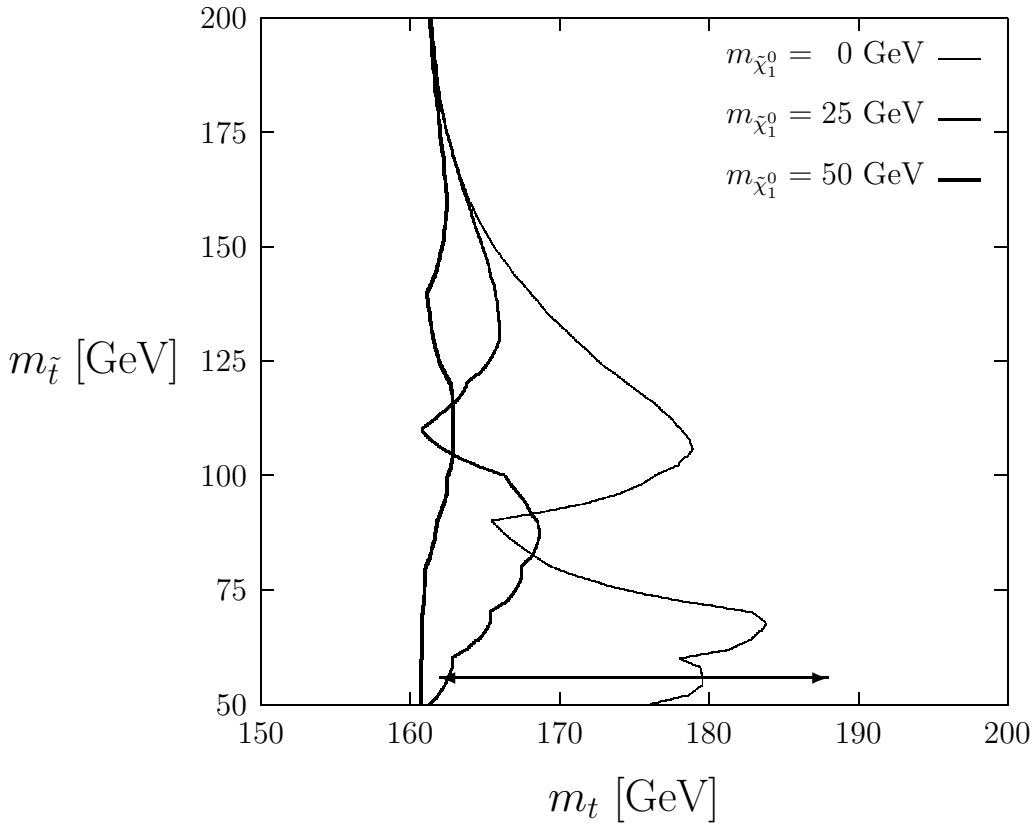

\begin{center}
\hspace*{6mm}
\setlength{\unitlength}{0.240900pt}
\ifx\plotpoint\undefined\newsavebox{\plotpoint}\fi
\sbox{\plotpoint}{\rule[-0.175pt]{0.350pt}{0.350pt}}%

\end{center}
\caption[]{Solutions of 
  $n(m_t,m_{\tilde t})=7-\sqrt{7}$ 
  ({\em cf.} Eq.~(\protect\ref{events})),
  for three different neutralino masses.
  The solutions of 
  $n(m_t,m_{\tilde t})=7+\sqrt{7}$
  are located much further to the left
  at too low top masses.
  The arrow indicates the $1\sigma$ top mass range from the CDF and 
  electro-weak measurements.}
\label{fig3}
\end{figure}

Equating the predicted number 
of dilepton and lepton+jet events Eq.~(\ref{events})
to the observed number of top candidate events (12)
minus the expected background (5)
modulo its Poisson fluctuation 
($7 \pm \sqrt{7}$),
we can establish contours of one standard deviation
in the $(m_t,m_{\tilde{t}})$ plane.  
These are shown in Fig.~\ref{fig3}
for three different values of the neutralino mass,
where the areas to the right of the curves are forbidden 
at this confidence level.
The theoretical uncertainty is added linearly,
{\em i.e.}, it is taken into account by using in Eq.~(\ref{events})
the upper bound of the cross sections displayed in Fig.~\ref{fig1}.
If the stop squark is heavy or absent
the rates observed by CDF are incompatible 
with a top mass in excess of 162 GeV at the $1\sigma$ level,
in contradiction with the mass 
inferred by CDF from the reconstruction of the 10 lepton+jet events, 
$174 \pm 10{}^{+13}_{-12}$ GeV combined with the electro-weak measurement 
$m_t = 178 \pm 11 {}^{+18}_{-19}$ GeV.
Combining these estimates of the top quark masses and the CDF 
cross section gives a minimum $\chi^2=1.6$ at $m_t = 164$ GeV.  
The $1\sigma$ range is indicated by the arrow on Fig.~\ref{fig3}.

It appears from Fig.~\ref{fig3}
that a light stop squark is not likely to solve this puzzle.
Only in the most optimistic scenario, 
when the neutralino is lighter than about 30 GeV,%
\footnote{A massless neutralino 
cannot be completely excluded if $\tan\beta<2$.} 
the stop squark mass is below 130 GeV 
and this mass is not within 10 GeV of the $\tilde{\chi}^0_1Wb$ threshold,
one obtains agreement with the mass measurements at the $1\sigma$ level (down 
from $1.6\sigma$).
One might argue 
that if stops are produced along with tops
the event reconstruction is bound to yield unreliable results.
However,
the number of stop events contributing is too small
to seriously invalidate the kinematical mass determination of the top quark.

The reason why the stop signal 
is hardly included in the CDF sample is simple.  
If the stop is heavy
(say $m_{\tilde t}>150$ GeV)
its production cross section is too low.
On the other hand,
if the stop is light
(say $m_{\tilde t}<150$ GeV)
the two non-observed neutralinos carry off so much energy, 
that the remaining visible particles 
have barely enough energy left over to pass 
the strong overall 85 GeV transverse momentum cut
imposed by CDF on the signal.
This can be clearly seen 
in Figs~\ref{fig2} and \ref{fig3},
where in the neighbourhood of the $\tilde{\chi}^0_1Wb$ thresholds
the stop efficiencies drop to zero.
Only almost massless neutralinos leave enough events to make a difference.


\section{Conclusion}

It has been suggested \cite{stop}
that the excessive top quark cross section 
observed by the CDF collaboration
can be explained by the production of light stop squarks,
whose decays mimic the top signals.
We have shown that this is unlikely to be the case.
Although the signals look the same superficially 
and the cross sections can be similar, 
some of the cuts cuts used by CDF to perform their analysis
exclude many possible stop events from the observed sample in
all but a small corner of the supersymmetric parameter space
($m_{\tilde{\chi}^0_1Wb} < 30$ GeV, $m_{\tilde{t}} < 130$ GeV and not within 
10 GeV from the $\tilde{\chi}^0_1Wb$ threshold).
Note, though, 
that this conclusion only concerns the observability of stop squarks 
in the top sample; 
the possibility of observation in dedicated searches is not touched upon.

\begin{ack}
It is a pleasure for F.C.~to thank
Wolfgang Ochs,
Stuart Samuel and
Leo Stodolski
for interesting discussions and comments.
K.J.A. would like to thank the Max-Planck-Institut f\"ur Physik 
for hospitality during the course of this work.
\end{ack}



\begin{thebibliography}{10}
\bibitem{CDFtop}
  CDF Collaboration,
  {\em Phys.~Rev.~Lett.} {\bf73} (1994) 225
  [hep-ex/9405005];
  {\em Phys.~Rev.~D} {\bf50} (1994) 2966-3026.
\bibitem{somerecentreview}
  B.~Jacobsen, 
  Rencontres de Moriond: QCD and High Energy Hadronic Interactions, 
  Meribel les Allues, France, 19-26 March 1994,
  CERN-PPE-94-97, Jun 1994
  [hep-ex/9407002];\\
  G.~Montagna, O.~Nicrosini, G.~Passarino, F.~Piccinini,
  HEPPH-9407246, Jul 1994
  [hep-ph/9407246];\\
  S.~Matsumoto, 
  KEK-TH-418, Nov 1994
  [hep-ph/9411388].
\bibitem{stop}
  T.~Kon, T.~Nonaka,
  {\em Phys.~Rev.~D} {\bf50} (1994) 6005
  [hep-ph/9405327];\\
  W.~de Boer, R.~Ehret, D.I.~Kazakov,
  {\em Phys.~Lett.~B} {\bf334} (1994) 220
  [hep-ph/9405419];\\
  J.L.~Lopez, D.V.~Nanopoulos, A.~Zichichi, 
  CERN-TH-7296-94, Jun 1994
  [hep-ph/9406254].
\bibitem{er}
  J.~Ellis, S.~Rudaz, 
  {\em Phys.~Lett.~B} {\bf128} (1983) 248.
\bibitem{susy}
  H.P.~Nilles, {\em Phys.~Rep.} {\bf 110} (1984) 1; \\
  H.E.~Haber and G.L.~Kane, {\em Phys.~Rep.} {\bf 117} (1985) 75.
\bibitem{DawsonEichtenQuiggSuper}
  S.~Dawson, E.~Eichten, C.~Quigg,
  {\em Phys.~Rev.~D} {\bf31} (1985) 1581.
\bibitem{GRV}
  M.~Gl\"uck, E.~Reya, A.~Vogt,
  {\em Z.~Phys.~C} {\bf53} (1992) 127.
\bibitem{JackWillyTop}
  E.~Laenen, J.~Smith, W.L.~van Neerven,
  {\em Phys.~Lett.~B} {\bf321} (1994) 254
  [hep-ph/9310233]
\end{thebibliography}
\end{document}